\documentclass[prb,twocolumn,superscriptaddress]{revtex4-1}
\usepackage{graphicx}
\usepackage{dcolumn}
\usepackage{bm}
\usepackage{color}
\newcommand{\tto}[1]{Tb$_2$Ti$_2$O$_7${#1}}
\newcommand{\hto}[1]{Ho$_2$Ti$_2$O$_7${#1}}
\newcommand{\dto}[1]{Dy$_2$Ti$_2$O$_7${#1}}

\newcommand{\ybto}[1]{Yb$_2$Ti$_2$O$_7${#1}}

\usepackage{amssymb}
\usepackage{epstopdf}

\begin{document}

\title{Sample independence of magnetoelastic excitations in the rare earth pyrochlore \tto{}}
\author{M Ruminy}
\affiliation{Laboratory for Neutron Scattering and Imaging, Paul Scherrer Institut, 5232 Villigen PSI, Switzerland}
\author{L Bovo}
\affiliation{London Centre for Nanotechnology and Department of Physics and Astronomy, University College London, 17-19 Gordon Street, London, WC1H 0AH, UK}
\author{E Pomjakushina}
\affiliation{Laboratory for Scientific Developments \& Novel Materials, Paul Scherrer Institut, 5232 Villigen PSI, Switzerland}
\author{M K Haas}
\altaffiliation{Now at Air Products and Chemicals Inc., Allentown PA 18195 USA}
\affiliation{Department of Chemistry, Princeton University,  Princeton NJ 08540, USA}
\author{U Stuhr}
\affiliation{Laboratory for Neutron Scattering and Imaging, Paul Scherrer Institut, 5232 Villigen PSI, Switzerland}
\author{A Cervellino}
\affiliation{Swiss Light Source, Paul Scherrer Institut, 5232 Villigen PSI, Switzerland}
\author{R J Cava}
\affiliation{Department of Chemistry, Princeton University,  Princeton NJ 08540, USA}
\author{M Kenzelmann}
\affiliation{Laboratory for Scientific Developments \& Novel Materials, Paul Scherrer Institut, 5232 Villigen PSI, Switzerland}
\author{T Fennell}
\email{tom.fennell@psi.ch}
\affiliation{Laboratory for Neutron Scattering and Imaging, Paul Scherrer Institut, 5232 Villigen PSI, Switzerland}

\date{\today}

\begin{abstract}

Recent experimental results have emphasized two aspects of \tto{} which have not been taken into account in previous attempts to construct theories of \tto{}: the role of small levels of structural disorder, which appears to control the formation of a long-range ordered state of as yet unknown nature; and the importance of strong coupling between spin and lattice degrees of freedom, which results in the hybridization of crystal field excitons and transverse acoustic phonons.  In this work we examine the juncture of these two phenomena and show that samples with strongly contrasting behavior vis-a-vis the structural disorder (i.e. with and without the transition to the ordered state), develop identical magnetoelastic coupling.  We also show that the comparison between single crystal and powder samples is more complicated than previously thought - the correlation between lattice parameter (as a measure of superstoichiometric Tb$^{3+}$) and the existence of a specific heat peak, as observed in powder samples, does not hold for single crystals.
\end{abstract}

\maketitle

\section{\label{sec:Introduction}Introduction}

The rare earth pyrochlores~\cite{Subramanian:1983vd} $R_2$Ti$_2$O$_7$ ($R=$ Gd-Yb) are well known as model systems for the study of frustrated magnetism~\cite{Gardner:2010fu}.  Although \tto{} has been studied as long and intensively as other members of the series such as the spin ices \dto{} and \hto{}, it remains a much less tractable mystery.  The low temperature state of \tto{} was originally thought to have only very short-range correlations amongst its spins~\cite{Gardner:1999ul}, which themselves continued to fluctuate down to the lowest temperatures measured~\cite{Gardner:1999ul,Gardner:2001kv,Gardner:2003wn}, often leading to the designation of \tto{} as a spin liquid.  Theories of \tto{}, which have aimed to explain the existence and properties of this state face a three-pronged problem: how does ostensibly unfrustrated \tto{} evade long-range magnetic order; why is there no cooperative Jahn-Teller transition, despite a non-Kramers doublet ground state; and does the state support any interesting emergent phenomena?  

Much theoretical progress has been made on the subject of spin liquids~\cite{Balents:2010jx,Normand:2009wk}, in particular in the area of quantum spin ices~\cite{Hermele:2004gg,Gingras:2014ip}.  The idea that \tto{} is closely related to spin ices, with an additional quantum ingredient, is a long-running thread in the study of \tto{}~\cite{Gingras:2000vd,Molavian:2007ve,Benton:2012ep,Curnoe:2013iz} and is a good reason for continued interest in it, particularly given recent progress in the development of generalized Hamiltonians for rare earth pyrochlores~\cite{Onoda:2010jf,Onoda:2011ct,Lee:2012dx,Ross:2011tv,Savary:2012cq,Applegate:2012ci,HanYan,Gingras:2014ip}.  However, there is currently no theoretical consensus on the low temperature state of \tto{}~\cite{Kao:2003ex,Enjalran:2004kr,Molavian:2007ve,Bonville:2011dw,Petit:2012ko,Sazonov:2013cp,Curnoe:2013iz,Curnoe:2008gy,Curnoe:2007hd,Klekovkina:2011vv,Jaubert:2015fm}, and indeed it may not be possible to form one while certain experimental issues remain unresolved.

\tto{} is known to have antiferromagnetic interactions ($\theta_{CW}\approx-13$ K), and a non-Kramers doublet groundstate which produces an Ising-like moment that points ``in'' or ``out'' of the tetrahedra.  This is the first aspect of the \tto{} puzzle - with antiferromagnetic coupling, such moments should order in the unfrustrated ``all-in-all-out'' state.  Although the crystal field scheme has been investigated several times, the exact composition of the doublet is still discussed~\cite{aleksandrov,Gardner:2001kv,Gingras:2000vd,Malkin:2004ke,Mirebeau:2007hi,Malkin:2010fe,Lummen:2008fh,Maczka:2008bd,Gaulin:2011ba,Bertin:2012gu,Zhang:2014ef,Klekovkina:hc,Princep:2015kt}.  A static distortion resulting in a single-ion singlet groundstate~\cite{MAMSUROVA:1986wx,Rule:2009wa,Chapuis:2010ir,Bonville:2011dw}, the second possibility in the puzzle, has never been accepted since the ground state apparently does retain a permanent magnetic moment on the terbium ions and no departure from cubic symmetry can be detected~\cite{Ruff:2007hf,Ofer:2007kt,Sazonov:2013cp}.  The original picture of very short-range spin correlations~\cite{Gardner:2001kv}; spin fluctuations on the timescale of probes such as $\mu$SR~\cite{Gardner:1999ul}, neutron spin echo and susceptibility\cite{Gardner:1999ul,Gardner:2003wn}; and a large quasielastic contribution to the neutron scattering~\cite{Yasui:2002cm,Mirebeau:2007hi,Rule:2009wa,Bonville:2011dw,Gaulin:2011ba,Takatsu:2011kg} has given way to a Coulomb phase with power-law correlations~\cite{Fennell:2012ci} and propagating excitations~\cite{Guitteny:2013hf}, and/or a mesoscopically ordered spin ice state~\cite{Fritsch:2013dv,Fritsch:2014cx,Guitteny:arx}.  The evolution and dynamics of this state are rather unclear.  The spin correlations of the spin liquid phase begin to build up below $T\sim40$ K and at first are isotropic and short-ranged~\cite{Gardner:2001kv}, becoming anisotropic and complex by $T=1.7$ K~\cite{Fennell:2012ci}.  An enhancement of the mesoscopic spin ice correlations has been observed below 0.275 K~\cite{Fritsch:2014cx} by neutron scattering, but other freezing transitions have also been found at temperatures of 0.1 K~\cite{Luo:2001wx}, 0.15 K~\cite{Yaouanc:2011be}, 0.2 K~\cite{Hamaguchi:2004dt,Lhotel:2012hb}, and 0.35 K~\cite{Gardner:2003wn} by techniques including $ac$-susceptibility and $\mu$SR.  Similarly, the presence or absence of a magnetization plateau with field applied along $[111]$, a feature whose existence~\cite{Molavian:2007ve,Molavian:2009dl} or absence~\cite{Sazonov:2013cp} could vindicate certain theories of \tto{}, has been debated~\cite{Legl:2012gm,Yin:2013iz,Sazonov:2013cp,Baker:2012cd,Lhotel:2012hb}.    

These last two points concerning the low temperature state, particularly the various temperatures for freezing transitions, suggest there is sample dependence in \tto{}.  In fact, this is most pronounced as the presence (absence) of a specific heat peak~\cite{Gardner:1999ul,Siddharthan:1999ww,Gingras:2000vd,Gardner:2003wn,Hamaguchi:2004dt,Chapuis:2010ir,Takatsu:2011kg,Yaouanc:2011be,Taniguchi:2013fi,Fennell:2014gf}, which is thought to be due to a transition to a long range ordered state of unknown character at $T\approx0.5$ K.  The sample dependence in \tto{} is sometimes suggested to be due to differences between single crystal and powder samples, with the former being variable and the latter reproducible~\cite{Taniguchi:2013fi}.  

A similar situation exists in \ybto{}, in which powder samples have extremely sharp heat capacity anomalies, while crystals have at best weak and broad anomalies~\cite{Ross:2012dj,Yaouanc:2011bj}.  This difference was suggested to be due to ``light stuffing'' of rare earth ions (i.e. substitution of Yb$^{3+}$ at the Ti$^{4+}$ site to give a stoichiometry like Yb$_{2+x}$Ti$_{2-x}$O$_y$) due to the evaporation of a small amount of titanium during the growth of a single crystal from nominally stoichiometric powders~\cite{Ross:2012dj}.  This process is attributed to the high temperatures used during crystal growth, and can only produce an excess of rare earth ions, since titanium is preferentially evaporated.  Because lower temperatures are employed during powder synthesis, the evaporation of titanium is not thought to be a problem, so that the stoichiometry of the starting materials is preserved, and either rare earth-depleted or rare earth-rich (stuffed) powder samples can be produced.  The role of this effect in \tto{} was studied by synthesizing powder samples of Tb$_{2+x}$Ti$_{2-x}$O$_{7-y}$, and a strong effect was indeed observed as a function of $x$~\cite{Taniguchi:2013fi}: the heat capacity peak was absent for small negative values, and suddenly appears around $x=-0.005$.  For larger $x$, including $x\approx0$, a strong heat capacity peak occurs at $T\approx0.5$ K.  It is accompanied by a very clear splitting of the quasielastic scattering into a new sharp mode at 0.1 meV, and a small, presumably magnetic, Bragg peak at $k=(1/2,1/2,1/2)$.  Further studies show that the mesoscopic correlations are present in all samples, but the resolution limited Bragg peak only occurs in association with the heat capacity peak~\cite{Guitteny:arx}.  Interestingly, the small size of the Bragg peak shows it cannot be due to the ordering of the full moment of the Tb$^{3+}$ ground state doublet, so the nature of the order parameter remains unknown.  Recently is has been suggested to be due to a quadrupolar transition~\cite{Kadowaki:2015_1,Kadowaki:2015_2}, but since the long range order associated with the heat capacity anomaly has so far not been directly determined, we will refer to it as a ``hidden order'' throughout the rest of the paper.  

There is also (at least) one more important factor in the physics of \tto{} which must be understood - the lattice.  It is well known that there is a strong spin lattice coupling in \tto{}, as manifested in elastic constants~\cite{MAMSUROVA:1986wx,MAMSUROVA:1988wg2,Nakanishi:2011bz}, dielectric constant~\cite{Mamsurova:1985ug}, anisotropic strain~\cite{Ruff:2007hf}, thermal conductivity~\cite{Li:2013ko} and thermal Hall effect~\cite{Hirschberger:2015he}, which all become anomalous exactly at the onset of the spin liquid regime at $10<T<40$ K.  Pressure induced magnetic order~\cite{Mirebeau:2002fm} and field induced lattice modifications~\cite{Ruff:2010uq} also indicate the coupling.  Most recently, the interaction was shown to give rise to hybridization between crystal field excitons and transverse acoustic phonons, forming a so-called magnetoelastic mode (MEM), suggesting that pure spin models cannot capture the physics of \tto{}, since the basic degrees of freedom of \tto{} would not be accurately incorporated~\cite{Guitteny:2013hf,Fennell:2014gf}.  The fate of the magnetoelastic coupling at the ordering transition is not known, though such a coupling can be very important in quadrupolar systems.  

In our previous study of the MEM, we used a single crystal with no heat capacity peak, implying a level of defects sufficient to destroy the hidden order state.  In the garnet Tb$_3$Ga$_5$O$_{12}$ (TGG), which also supports a thermal Hall effect~\cite{Strohm:2005gs}, superstoichiometric terbium ions have recently been advanced as a source of magnetic phonon scattering underlying the effect~\cite{Mori:2014fe}, and one might therefore ask if the spin-phonon interaction we observed in \tto{} is actually due to disorder.   In this work, we compare crystals with (without) the specific heat anomaly, implying, by the logic of Ref.~\onlinecite{Taniguchi:2013fi} the absence (presence) of structural disorder, and show that in the range $1.5<T<50$ K all the crystals develop the same magnetoelastic excitation spectrum. Although we suggest that the various low temperature behaviors of \tto{} all emerge from the same microscopic degrees of freedom, which are formed by the magnetoelastic hybridization, we do not directly investigate their behavior at low temperature in this work. We also find that the comparison of the single crystal sample compositions with the study of the composition dependent properties~\cite{Taniguchi:2013fi} is not as simple as we expected, suggesting that other structural effects may be involved in controlling the ordering transition.

\section{\label{sec:Experimental_methods}Experimental methods}

\subsection{Crystal growth}

We discuss three different single crystals.  Sample MH1 is our original sample, grown at Princeton and discussed in Refs.~\onlinecite{Fennell:2012ci} and \onlinecite{Fennell:2014gf}.  Samples EP2 and EP3 were grown at PSI, and have not previously been reported.

As grown \tto{} crystals contain dark/black regions, and annealing of \tto{} crystals under oxygen converts them to a state which is uniform black/dark in color.  This is thought to be due to the formation of Tb$^{4+}$ defects, as in Pr$_2$Zr$_2$O$_7$ where transparent green crystals are thought to be ideal and brown/black crystals to be contaminated with Pr$^{4+}$.  Annealing under argon converts both as-grown or oxygen annealed \tto{} crystals to a uniform transparent reddish-orange.  This may be contrasted with pyrochlores such as \hto{} or \dto{}, where annealing under oxygen is used to produce transparent crystals of characteristic colors~\cite{Prabhakaran:2011hl}.  The exact conditions of growth for sample MH1 are not known, but prior to neutron scattering experiments it was annealed under flowing argon to eliminate dark patches and produce a large reddish-orange boule.  

For the growth of sample EP2 and EP3, polycrystalline \tto{} was prepared by a solid state reaction. Starting materials of Tb$_4$O$_7$ and TiO$_2$ with 99.99\% purity were mixed and ground, followed by heat treatment at $900-1150$ $^{\mathrm{o}}$C in air, for at least 100 hours with several intermediate grindings.  The phase purity of the resulting powder was confirmed using a conventional laboratory powder x-ray diffractometer.   This material was hydrostatically pressed in the form of rods of 7 mm diameter and $\approx60$ mm length. The rods were subsequently sintered at 1150 $^{\mathrm{o}}$C for 15 hours. The crystal growth was carried out using an optical floating zone furnace (FZ-T-10000-H-IV-VP-PC, Crystal System Corp., Japan) with four 1000 W halogen lamps as a heat source. The growth rate was 10 mm hr$^{-1}$, with both rods (feed and seed rod) rotated at 25 rpm in opposite directions to ensure homogeneity of the melt.  During growth, 2.5 bar pressure of argon-oxygen mixture (50:50) was applied. The obtained crystals were post-annealed for 48 hours at 1150 $^{\mathrm{o}}$C in argon.

\subsection{Heat capacity}

The specific heat of small pieces of each crystal was measured between 0.35 K and 50 K with a Quantum Design Physical Properties Measurement System (PPMS), equipped with a $^3$He option, using a heat-relaxation method. An addenda measurement was made to evaluate the background of Apiezon Grease N and this contribution was subtracted from the data.  Differing lattice contributions in different pyrochlores make an accurate estimation of magnetic contributions to the specific heat above $T\sim 10$ K difficult to evaluate.

\subsection{Neutron scattering}

The neutron scattering experiments performed on sample MH1 to investigate the magnetoelastic mode were described in Ref.~\onlinecite{Fennell:2014gf}, and involved both time of flight and triple axis spectroscopy.  We also report some new measurements in which the crystal, held in a copper clamp holder, was mounted in a dilution refrigerator and cryomagnet.  To confirm the existence, dispersion and temperature dependence of the magnetoelastic mode in samples EP2 and EP3 we used the thermal neutron triple axis spectrometer EIGER at SINQ, PSI.  The crystals, which were aligned such that the scattering plane contains wavevectors of type $(h,h,l)$, were mounted on aluminium holders in a standard helium cryostat.  We used a PG002 monochromator, analyzer and filter, and operated EIGER with fixed $E_f=14.68$ meV (i.e. $k_f=2.662$ \AA$^{-1}$).  The magnetoelastic mode can be readily located by constant energy scans in the otherwise featureless part of the spectrum between the intense first and second crystal field excitations (i.e. at energies of 4 - 7 meV), and in constant wavevector scans at positions such as $(1.5,1.5,0)$.

\subsection{X-ray diffraction}

Fragments from each crystal were mixed with silicon powder and ground together in an agate pestle and mortar to obtain uniform powders.  These mixtures were loaded into 0.3 mm glass capillaries.  The silicon serves primarily to disperse the \tto{} in the beam, while minimizing absorption, but also provides a convenient calibrant for wavelength and lattice parameters ($a_\mathrm{Si}=5.431194$ \AA ~at $22.5$ $^\mathrm{o}$C, NIST powder diffraction standard 640c).  We measured the diffraction pattern of the mixture using the high resolution powder diffractometer of the Materials Science Beamline at the Swiss Light Source (SLS).  The diffractometer operates in Debye-Scherrer geometry, using a Mythen microstrip detector, capillary spinner, and $2\theta$ range extending from $2^\mathrm{\circ}$ to 120$^\mathrm{\circ}$.  Sample MH1 was measured in a previous experiment in which the incident wavelength was $\lambda=0.620474(3)$ \AA~(i.e. $E=19.98$ keV), and sample EP2 and EP3 were measured in a second experiment using a wavelength of $\lambda=0.4959$ \AA~(i.e. $E=25$ keV).  All measurements were made at room temperature, which is maintained constantly at 24 $\deg$ C at the SLS.  The powder diffraction data were modeled and fitted using the Rietveld method, as implemented in the package FullProf~\cite{Rodriguez}.  

Details of the refinements for sample MH1 were presented in Ref.~\onlinecite{Fennell:2014gf}.  Notably, we found that while the shape of the Bragg peaks due to the silicon was well modeled by a pseudo-Voigt form, as expected for this diffractometer, the Bragg peaks of \tto{} were best described by a pure Lorentzian.  This effect is somewhat less pronounced with a shorter wavelength, but nonetheless, a pure Lorentzian lineshape gives the best description for the \tto{} samples (the difference being that the FWHM of the pseudo-Voigt becomes negative in certain angular ranges, while for the Lorentzian it does not).  We used a conventional Rietveld refinement of a crystallographic model incorporating two phases (i.e. \tto{} and silicon).  In general, we refined linearly interpolated background, profile parameters, and isotropic thermal parameters for both phases.  For \tto{}, we also refined the free positional parameter of the $48f$ oxygen site, and the lattice parameter.    The lattice parameter of the silicon standard (defined at $22.5$ $\deg$ C) was corrected for thermal expansion~\cite{Bergamin:1997co} at the temperature of the experimental hutch (24 $\deg$ C) and then held fixed, while the wavelength and zero-shift of $2\theta$ were refined.  Asymmetry corrections were applied up to $2\theta=25^\mathrm{o}$ and refined, but capillary offset parameters could not be stably refined.  All models were refined freely to convergence.

\section{\label{sec:Results}Results}

\begin{figure}
	\centering
	\includegraphics[trim=115 100 100 200,clip=true,scale=0.5]{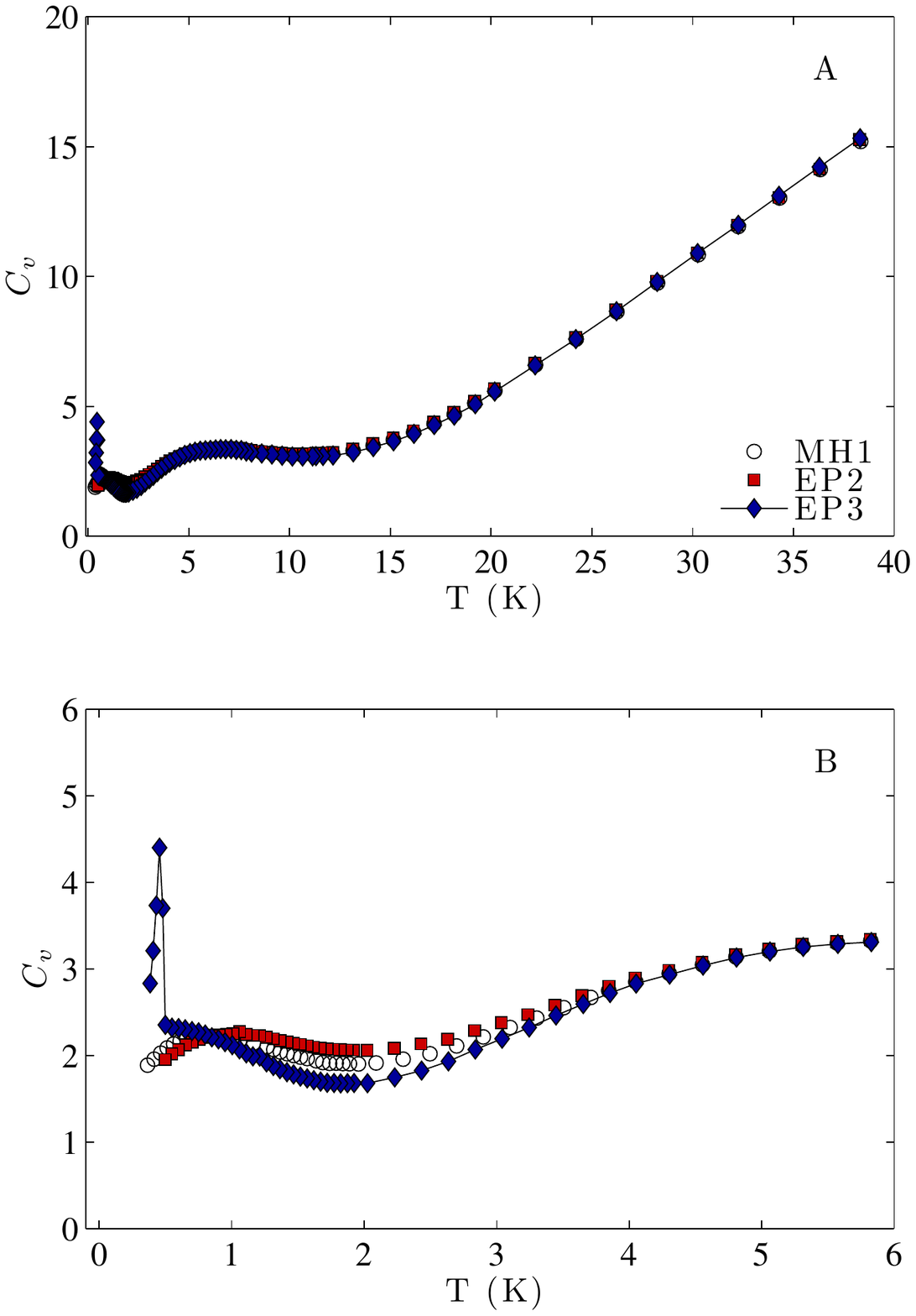}\\
	\caption{Comparison of the specific heat of the three crystals of \tto{}.  At temperatures above 4 K, the heat capacities are identical and show no feature associated with the onset of magnetoelastic coupling at $T\sim20$ K (panel A).  At low temperatures, samples MH1 and EP2 are similar, with a downturn below 1 K, while sample EP3 has a pronounced peak at $T\sim 0.5$ K (panel B). (The line for sample EP3 is a guide for the eye to highlight the sharp onset of the heat capacity peak.)}
	\label{fig:cv}
\end{figure}

The heat capacities of the three samples are identical down to 4 K, as shown in Figs.~\ref{fig:cv}A and \ref{fig:cv}B.  At temperatures in the range $10<T<40$ K where the magnetoelastic coupling develops, the heat capacities show that there is no sign of any symmetry breaking phase transition associated with the coupling.  At low temperatures, below 2 K, sample MH1 and EP2 exhibit the very broad peak at $T\approx1$ K which is typically attributed to the formation of the so-called spin liquid state in \tto{}.  Sample EP3 shows a more pronounced dip in the heat capacity at $T\approx2$ K, and the broad peak which appears below this suddenly gives way to a very sharp peak in the heat capacity at $T\approx0.46$ K.  This sharp feature is very similar to those observed in powder samples of Tb$_{2+x}$Ti$_{2-x}$O$_y$ with $0<x<0.005$~\cite{Taniguchi:2013fi}, where the temperature, sharpness and height of the peak are maximum for $x=0.005$.  By comparison, the height of the peak in sample EP3 suggests $0.0025<x<0.005$.

\begin{figure}
	\centering
	\includegraphics[trim=100 140 100 100,clip=true,scale=0.5]{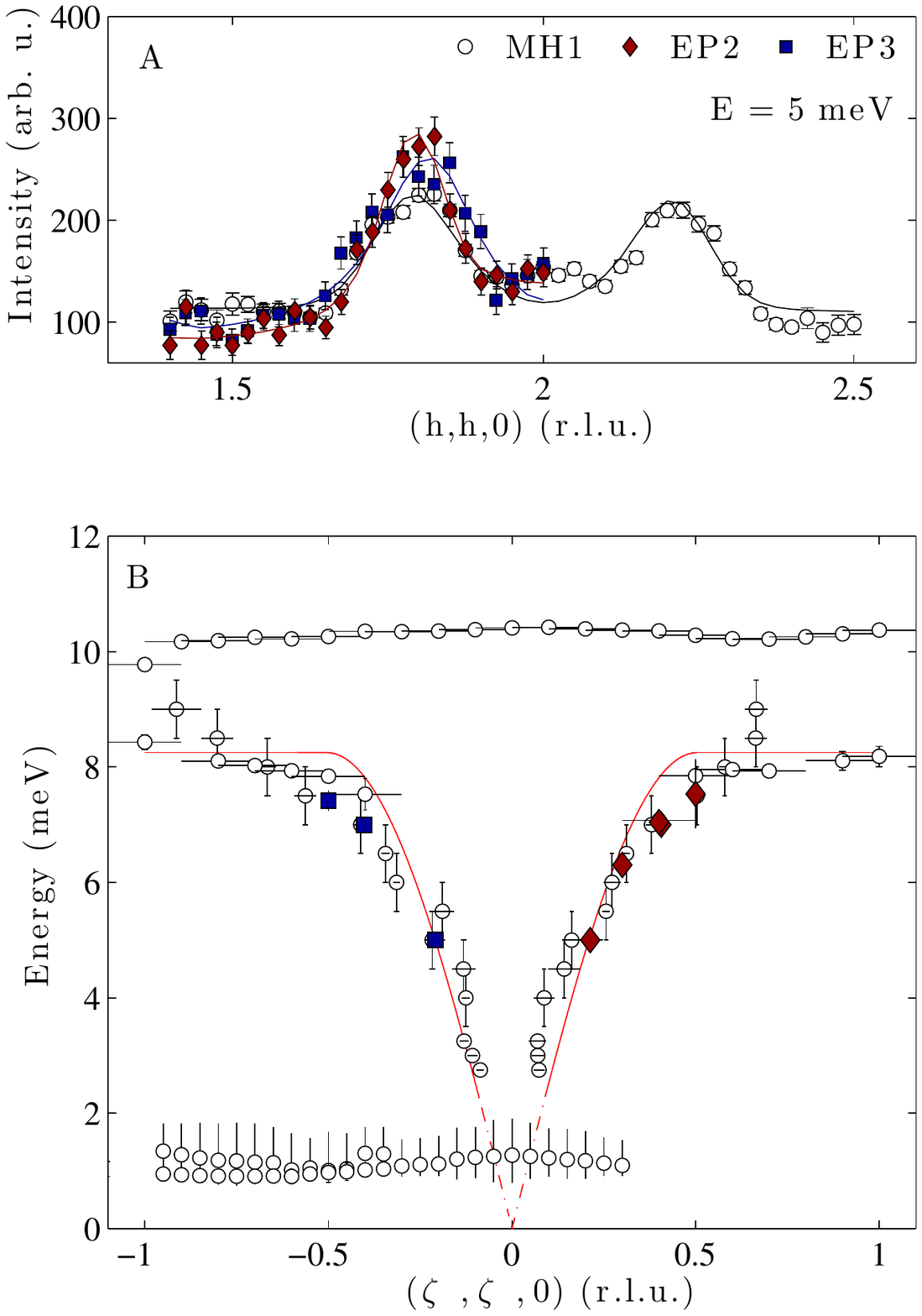}\\
	\caption{Existence and dispersion of MEMs in three crystals of \tto{}.  The mode is found in all the crystals using longitudinal constant energy scans along the $(h,h,0)$ axis at $(2,2,0)$ with energy transfer $E=5$ meV (panel A).  Constant energy scans ($E=5,7$ meV) and constant wavevector ($\zeta=-0.5$) show that the dispersion relation of the MEM is the same in all three crystals (panel B).  (Data for MH1 shows the full structure of the excitation spectrum, including the MEM and first and second crystal field excitons, as already presented in Ref.~\onlinecite{Fennell:2014gf}, EP2 peak positions are reflected to positive $\zeta$ for clarity.  The line in B is a guide to the eye for the dispersion relation of the form $\hbar\omega=a\sim{(0.7|k|\pi)}$ for $-0.7<|k|<0.7$ and $\hbar\omega=a$ for $\pm(0.7<k<1)$. $a=8.25$ meV, the approximate zone boundary energy, and the crossover to a wavevector-independent section is due to the fact that at $\zeta=\pm0.75$, the plotted dispersion relation runs along the Brillouin zone boundary.)}
	\label{fig:memdisps}
\end{figure}

\begin{figure}
	\includegraphics[trim=100 500 100 75,clip=true,scale=0.5]{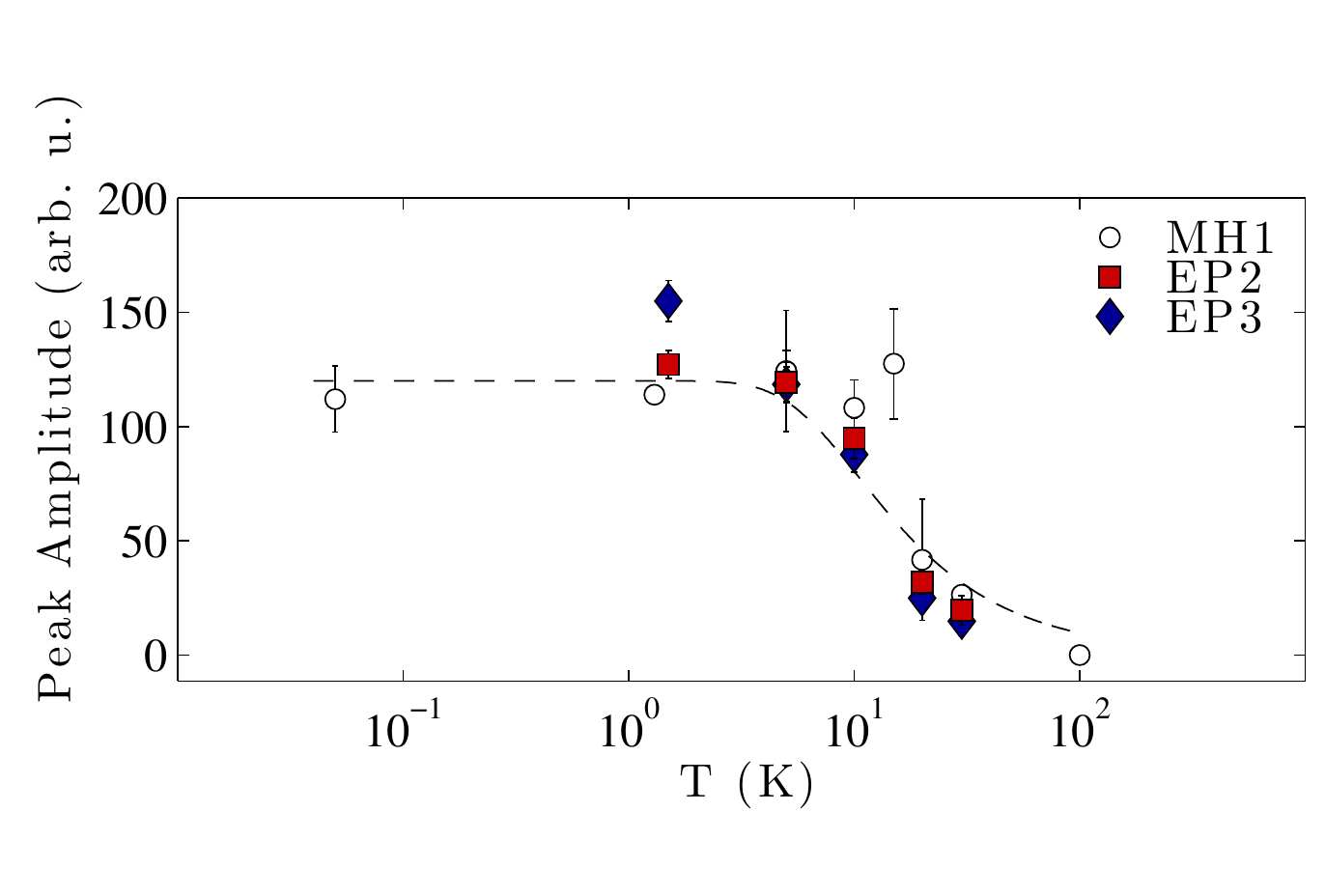}
	\caption{Temperature dependence of magnetoelastic mode intensitites in three crystals of \tto{}.  (Intensities from different crystals are scaled to match at 5 K.  The dashed line is $n_0-n_1$ (scaled), where $n_0$ and $n_1$ are the thermal population factors of the ground and excited states, respectively, of a two-level system with $\Delta=1.4$ meV.)}
	\label{fig:memtdep}
\end{figure}

On cooling, the hybrid excitations develop in the temperature range $10<T<40$ K, and then remain essentially unchanged to temperatures as low as 0.05 K~\cite{Fennell:2014gf}.  We can therefore conveniently investigate their existence without recourse to low temperature sample environment equipment.  In Fig.~\ref{fig:memdisps}A we show constant energy scans along the $(h,h,0)$ direction at $(2,2,0)$ with an energy transfer of 5 meV, at 1.5 K.  We see that in all three crystals, a steeply dispersing, longitudinal excitation exists (i.e. as we scan along $(h,h,0)$ we cut through the dispersion surface twice, resulting in a double peak).  That the dispersion is the same in all the samples is confirmed in Fig.~\ref{fig:memdisps}B where we show the full structure of the excitation spectrum as it was presented in Ref.~\onlinecite{Fennell:2014gf}, along with a limited number of points obtained from various scans through the excitations of the new samples.  We can also follow the temperature dependence of the hybrid modes, and show in Fig.~\ref{fig:memtdep} that the three samples are essentially identical in this respect.  The comparison of heat capacity and neutron scattering experiments shows directly that the hybrid modes are robust against the sample dependence which affects the low temperature state.  

\begin{figure}
	\centering
	\includegraphics[trim=60 465 50 90,clip=true,scale=0.5]{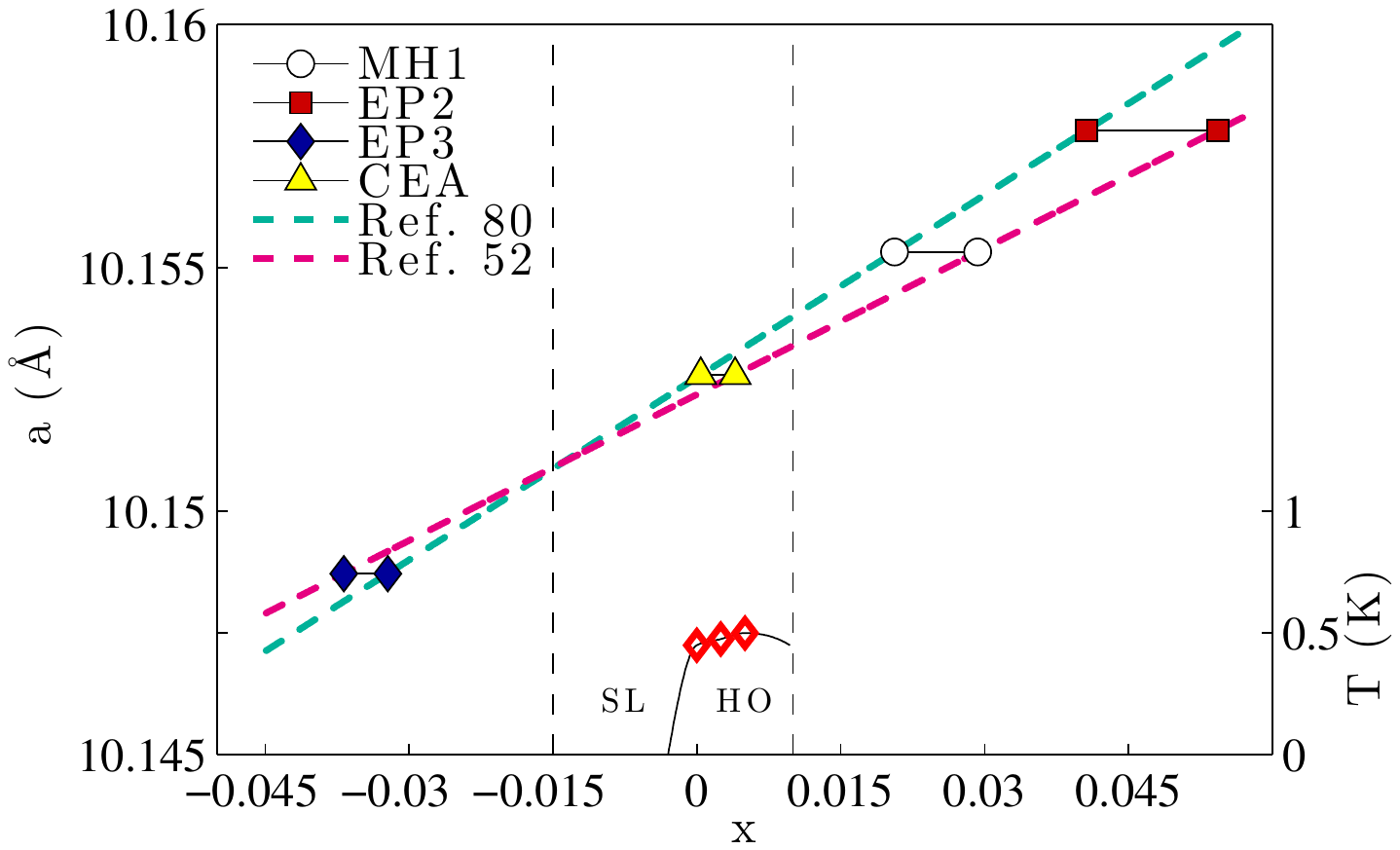}\\
	\caption{Lattice parameter trends and stoichiometry estimates for various \tto{} crystals.  The measured lattice parameters are used to obtain the composition by comparison with lattice parameter trends~\cite{Taniguchi:2013fi,Taniguchi:2015,Guitteny:arx}.  Although all the crystals are nominally stoichiometric ($x=0$), their lattice parameters imply a considerable range of compositions, so the lattice parameter trends have been extrapolated as required.  The MEM, which was originally reported in sample MH1 in Ref.~\onlinecite{Fennell:2014gf}, and is here reported in EP2 and EP3, is also clearly visible in the data of Ref.~\onlinecite{Guitteny:2013hf}, measured using sample ``CEA''.  The vertical dashed lines indicate the window of stoichiometry studied in powders in Ref.~\onlinecite{Taniguchi:2013fi}, and the phase diagram for spin liquid (SL) and hidden order (HO) found in this window is also shown.  The composition and temperature dependence of the heat capacity peaks observed in Ref.~\onlinecite{Taniguchi:2013fi} are indicated by open (red) diamonds, and can be read from the bottom and right axes respectively.}
	\label{fig:a_vs_x}
\end{figure}

Given the strong connection between stoichiometry and form of heat capacity advanced in the work of Taniguchi {\it et al.}~\cite{Taniguchi:2013fi}, and the expectation that the stoichiometry of a pyrochlore can be diagnosed by accurate lattice parameter measurements~\cite{Taniguchi:2013fi,Fennell:2014gf,Ross:2012dj}, we use the lattice parameters obtained from  the x-ray diffraction in this way.  In Ref.~\onlinecite{Fennell:2014gf} we used the $x$-dependence of the lattice parameter of Tb$_{2+x}$Ti$_{2-x}$O$_y$ reported in Ref.~\onlinecite{Taniguchi:2013fi} to establish the stoichiometry of sample MH1 (Tb$_{2.013}$Ti$_{1.987}$O$_{6.99}$).  Following a correction of the lattice parameter calibration~\cite{Taniguchi:2015} and correction for thermal expansion of the silicon standard, we now find that sample MH1 ($a=10.15533(1)$ \AA) is Tb$_{2.02}$Ti$_{1.98}$O$_{6.99}$, sample EP2 ($a=10.15782(1)$ \AA) is Tb$_{2.04}$Ti$_{1.96}$O$_{6.98}$, and sample EP3 ($a=10.14873(1)$ \AA) is Tb$_{1.97}$Ti$_{2.03}$O$_{7.035}$.  The MEM also exists in the data presented in Ref.~\onlinecite{Guitteny:2013hf}, where the lattice parameter is reported to be $a=10.1528(5)$ \AA, suggesting a composition of essentially \tto{} (i.e. $x=0.0004$).  Fig.~\ref{fig:a_vs_x} shows the extrapolated lattice parameter trend of Ref.~\onlinecite{Taniguchi:2015} and indicates how these samples fall on it.  A new lattice parameter trend was recently reported, which we also show, and by comparison with this our samples would have composition Tb$_{2.029}$Ti$_{1.971}$O$_{6.986}$ (MH1), Tb$_{2.054}$Ti$_{1.946}$O$_{6.973}$ (EP2), and Tb$_{1.963}$Ti$_{2.037}$O$_{7.019}$ (EP3) respectively.   The lattice parameters of the nominally stoichiometric powder used for the crystal growth of samples EP2 and EP3, as well as a separate powder sample prepared by the same method, are clustered in the range $a=10.1525(4)$.  Although not shown on Fig.~\ref{fig:a_vs_x}., they all fall at $x\approx0$.

Fig.~\ref{fig:a_vs_x} also shows the window of compositions investigated in Ref.~\onlinecite{Taniguchi:2013fi}, and the proposed phase diagram of the hidden order.  We see that there appears to be considerable variation in the lattice parameter of nominally stoichiometric single crystals, far outside the window investigated in powder samples.  Furthermore, the sample which shows the heat capacity anomaly (EP3) does not fall within the window of stoichiometry expected by comparison with the powder samples (either in terms of lattice parameter or, as mentioned above, the height of the specific heat peak).  

\begin{figure}
	\centering
	\includegraphics[trim=1 1 1 1,clip=true,scale=0.5]{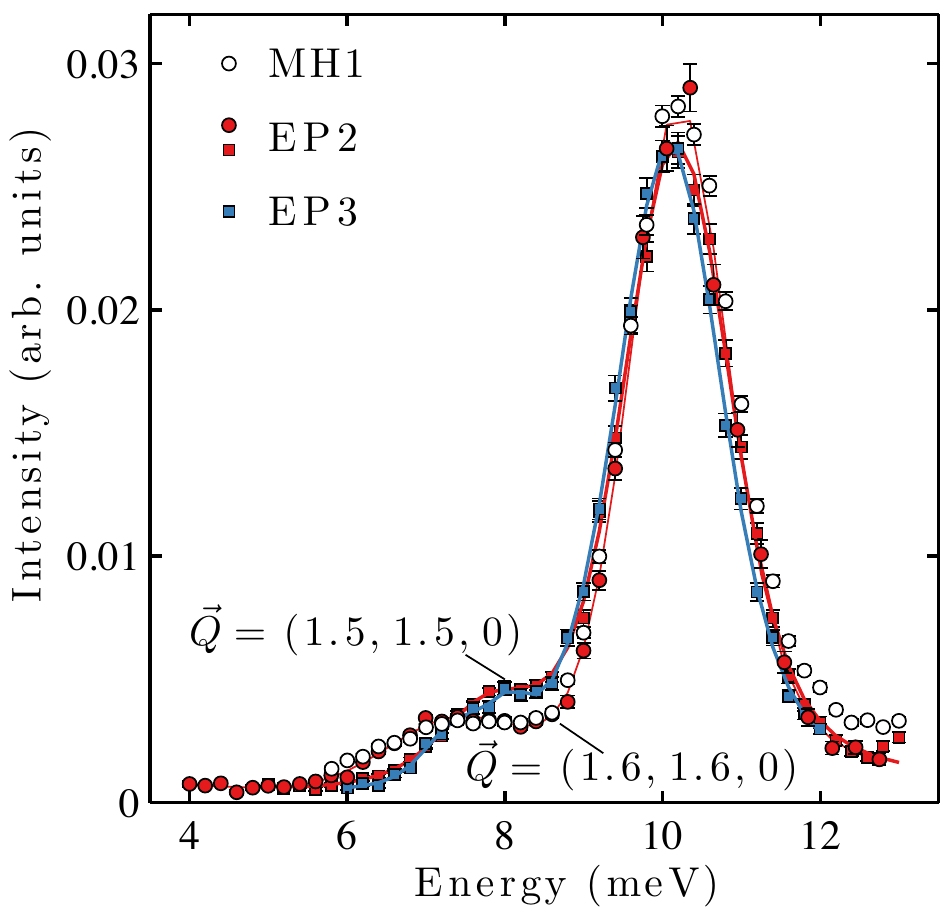}\\
	\caption{Comparison of the intensity of MEM and crystal field excitations for different crystals. When the signals from the intense crystal field excitation at 10.2 meV in the different crystals are scaled together, we find that the intensity of the MEM also scales exactly.}
	\label{fig:vol_fract}
\end{figure}

Another question related to the stoichiometry of the sample is the homogeneity.  In Fig.~\ref{fig:vol_fract} we compare the intensity of the MEM with the intensity of the second crystal field excitation (i.e. the mode at $10.2$ meV in Fig.~\ref{fig:memdisps}b) in constant wavevector scans.  We have measured such scans at $(1.6,1.6,0)$ in MH1 and EP2, and at $(1.5,1.5,0)$ in EP2 and EP3.  The measurements of MH1 were made at 0.07 K, while those of EP2 and EP3 were at 1.5 K, but it can be seen from Fig.~\ref{fig:memtdep} that the intensity of the MEM does not change below 10 K.  In both cases, when the peaks from the crystal field excitation are scaled together, the MEM also scales.  This result suggests that both excitations exist in the same volume fraction of all the crystals.  The background scattering in these experiments seems to come mainly from the sample itself, so the almost identical signal to noise ratio seen in Fig.~\ref{fig:memdisps}a is also to be expected in the case that the MEM exists throughout the sample.  However, comparison with another signal originating uniquely from the sample, as in Fig.~\ref{fig:vol_fract}, avoids any complications in this comparison related to different sample environment or sample holders.

\section{\label{sec:Discussion}Discussion}

We discuss first the qualitatively simple results obtained by comparing the heat capacity and neutron scattering experiments, and then turn to the thornier question of sample composition as quantified by x-ray diffraction. 

We have observed that the magnetoelastic coupling which results in the hybridization of crystal field excitons and transverse acoustic phonons in \tto{} is an intrinsic feature, which does not depend on the sample - crystals with and without the heat capacity anomaly have the same magnetoelastic excitation spectrum.  This may not seem surprising, since the hybridization develops at a much higher temperature than the heat capacity features attributed to the formation of the spin-liquid state or transition to hidden order.  However, these low-temperature phenomena must derive from low-energy degrees of freedom which emerge in a state whose basic structure always incorporates the magneotelastic coupling, and this is the first important aspect of our observations.  Secondly, while the presence (absence) of a heat capacity peak is associated with the absence (presence) of structural disorder, we show that the magnetoelastic coupling is not a consequence of such disorder.  Although superstoichiometric terbium ions in TGG have been advanced as the origin of certain magneto-phonon interactions~\cite{Mori:2014fe}, if structural disorder is indeed responsible for determining the low temperature state of \tto{} samples, it does not mediate the magnetoelastic coupling.  It will be very interesting to examine the fate of the magnetoelastic excitations at the ordering transition in samples such as EP3.

The remaining question is to understand the difference between samples MH1, EP2, and EP3.  Our first observation is that according to their lattice parameters, none of our nominally stoichiometric samples lie within the small compositional window around $x=0$ where the heat capacity peak is expected  (though the the lattice parameter reported in Ref.~\onlinecite{Guitteny:2013hf} places that sample within the window).

Although all the crystals are nominally stoichiometric, we see considerable variation amongst them, even between EP2 and EP3 which were grown by the same method in the same laboratory (and whose lattice parameters were measured consecutively in the same x-ray diffraction experiment, and which were synthesized from powders verified to have lattice parameters agreeing within $1\times10^{-4}$).  Other values of the lattice parameter of nominally stoichiometric \tto{} crystals or powders can be found in the literature clustered around 10.154 \AA~\cite{Subramanian:1983vd,Han:2004bz}, and also some which actually lie outside the range of Fig.~\ref{fig:a_vs_x}, such as 10.12 \AA~\cite{Rule:2009wa} or 10.1694 \AA~\cite{Lau:2006uc}.  At face value, this implies a surprisingly large range of off-stoichiometry, even amongst powders where the evaporation of titanium during synthesis is not thought to be problematic.  

Our second surprising observation is that the lattice parameter comparison implies a negative value of $x$ for sample EP3, i.e.terbium depletion during growth.  This is not compatible with the light stuffing mechanism, which depends on the evaporation of titanium during crystal growth, so can only produce superstoichiometric rare earth ions.  Rare earth depletion during crystal growth is not possible by this mechanism (as mentioned above, rare earth depletion is only possible in a powder sample by control of starting material stoichiometry and lower synthesis temperatures).  Because of the incompatibility of rare earth depletion and stuffing during crystal growth, factors other than this must be at play in order to have ``$x<0$'' in a single crystal, and the change of lattice parameter between starting material and crystal implies that they are associated with the growth process.  For example, oxygen deficient defect clusters have recently been detected in \dto{}~\cite{Sala:2014kz}, and, if present in \tto{} crystals in variable density, could perhaps modify the lattice parameter differently to the stuffing.  It will be interesting to compare the microstructure of crystals with and without the heat capacity anomaly.

Overall, we get the impression that determining stoichiometry of single crystals by comparison with published lattice parameter trends is more complicated than we had proposed in Ref.~\onlinecite{Fennell:2014gf}, which is now born out by the existence of two different lattice parameter trends in the literature~\cite{Taniguchi:2013fi,Taniguchi:2015,Guitteny:arx}.  Comparing lattice parameters measured under different experimental conditions may be more complicated than first suggested, requiring exact specification of the temperature and accurate wavelength calibration, which may not be possible retrospectively.  However, these effects are taken into account in the comparison of our crystals with the known lattice parameter trends.  

Very recently, a study was reported of a single crystal in which regions with different concentration of defects (covering essentially the full phase diagram of Ref.~\onlinecite{Taniguchi:2013fi}) could be identified by measuring the lattice parameter and specific heat of many small pieces cut along the length of the boule~\cite{Wakita:2015vr}.  It was suggested that large single crystals studied by neutron scattering may not be homogeneous.  Although our heat capacity samples were cut from our crystals adjacent to the (much larger) pieces used for neutron scattering experiments, there exists the possibility that they are inhomogeneous.  

We first note that the single crystal described in Ref.~\onlinecite{Wakita:2015vr} has a strong color gradient from red-orange to transparent accompanying the concentration gradient, while our crystals are each uniformly colored.  Secondly, in Fig.~\ref{fig:vol_fract}, we showed that the volume fraction of the crystal which supports the MEM is the same as that supporting the crystal field excitation.  An oft-cited advantage of neutron scattering is its sensitivity to the full volume of large samples, and since the crystal field spectrum is a universal property of all \tto{} samples~\cite{Lummen:2008fh}, we think it is justifiable to assume that the intensity of the crystal field excitation derives from the entire sample, and by virtue of its identical volume fraction, so does the MEM.  The crystal field excitation betrays no sign of sample dependence or inhomogeneity: in all the samples it has the same energy, identical Voigt peak shape, and identical width (1.7$\pm0.2$ meV, close to the estimated resolution limit ($1.3-1.5$ meV) of the spectrometer).  The MEM then is a property of \tto{}, robust to the levels of off-stoichiometry currently discussed.

More generally, we point out that single crystals studied by neutron scattering are highly similar, so far as data in the literature from different experiments can be compared.  For example, although different studies have employed different energy resolution/integration, or different wavevector resolution/detail, or polarization analyses, the diffuse scattering measured in crystals from four different groups appears to be quite compatible~\cite{Yasui:2002cm,Fennell:2012ci,Petit:2012ko,Fritsch:2013dv,Fritsch:2014cx}, and has recently also been shown not to depend on the form of the specific heat~\cite{kermarrec}.  Studies of the form of the dispersion of the first crystal field exciton report identical structures~\cite{Gardner:2001kv,Yasui:2002cm,Rule:2006fr,Guitteny:2013hf,Fennell:2014gf}.  Similarly, while polarization analysis and high resolution were employed to show that the quasielastic contribution contains a propagating mode~\cite{Guitteny:2013hf}, other unpolarized studies measure the total, which again is quite comparable~\cite{Yasui:2002cm,Yaouanc:2011be,Fennell:2012ci}.  Sadly the heat capacities of all these samples have not been reported, and this makes it extremely interesting to pursue studies of samples explicitly shown to have the heat capacity anomaly, in order to establish the fate of all these features at the transition.  Even if the true groundstate of pure \tto{} is a quadrupolar ordered state, there remain the questions of how a strongly correlated but disordered phase survives in the dipole sector, and, given that the ordered phase is destroyed by very small levels of disorder, does it support any of the interesting types of effect which may appear when a manifold of frustrated groundstates is perturbed by small levels of disorder~\cite{Sen:2011jt,Sen:td}? 

\section{\label{sec:Conclusion}Conclusion}

We have shown that the contrasting types of low temperature ($T<0.6$ K) states observed in different single crystals of \tto{}  - both hidden order and spin liquid - emerge from a higher temperature ($1.5 < T \lesssim 40$ K) state in which the same magnetoelastic excitation spectrum develops in all the samples.  We have shown that the comparison between single crystal samples and powders implies that other structural effects in addition to stuffing of terbium at the titanium site may also be involved in controlling the lattice parameter and eventual ordering behavior of the sample.
\\
\acknowledgements{
We acknowledge discussions with I Mirebeau, S Petit, S Guitteny, B D Gaulin, E Kermarrec, B Z Malkin, and H Kadowaki.  Neutron scattering experiments were carried out at the continuous spallation neutron source SINQ at the Paul Scherrer Institut at Villigen PSI in Switzerland, and work at PSI was partly funded by the Swiss NSF (grant 200021\_140862, Quantum Frustration in Model Magnets).  LB is supported by The Leverhulme Trust through the Early Career Fellowship program.}


%

\end{document}